\documentclass[aps,prd,preprint,superscriptaddress,nofootinbib]{revtex4-1}
\usepackage{epsfig,amssymb,bm,graphics,color}
\usepackage{epstopdf}
\usepackage{dcolumn}
\usepackage{amsmath}
\usepackage{bm}
\usepackage{slashed}
\usepackage[english]{babel}
\usepackage{siunitx}
\newcommand{\diff}{\mathrm d}
\renewcommand{\vec}[1]{\mathbf{#1}}

\begin{document}

\title{Laser assisted Breit-Wheeler and Schwinger processes}
\author{T.~Nousch}
\author{A.~Otto}
\affiliation{Institute of Radiation Physics, Helmholtz-Zentrum
Dresden-Rossendorf,\\ Bautzner Landstra\ss e 400, 01328 Dresden, Germany}
\affiliation{Institut f\"ur Theoretische Physik, Technische Universit\"at
Dresden,\\ Zellescher Weg 17, 01062 Dresden, Germany}
\author{D.~Seipt}
\affiliation{Helmholtz-Institut Jena, Fr\"obelstieg 3, 07743 Jena, Germany}
\affiliation{Theoretisch-Physikalisches Institut,
Friedrich-Schiller-Universit\"at Jena,\\ Max-Wien-Platz 1, 07743 Jena, Germany}
\author{B.~K\"ampfer}
\affiliation{Institute of Radiation Physics, Helmholtz-Zentrum
Dresden-Rossendorf,\\ Bautzner Landstra\ss e 400, 01328 Dresden, Germany}
\affiliation{Institut f\"ur Theoretische Physik, Technische Universit\"at
Dresden,\\ Zellescher Weg 17, 01062 Dresden, Germany}
\author{A.~I.~Titov}
\affiliation{Bogoliubov Laboratory for Theoretical Physics, JINR Dubna,\\
Joliot-Curie str.\ 6, 141980 Dubna, Russia}
\author{D.~Blaschke}
\affiliation{Bogoliubov Laboratory for Theoretical Physics, JINR Dubna,\\
Joliot-Curie str.\ 6, 141980 Dubna, Russia}
\affiliation{Institute for Theoretical Physics, University of Wroclaw,
pl.\ M.\ Borna 9, 50-204 Wroclaw, Poland}
\affiliation{National Research Nuclear University (MEPhI), Kashirskoe
Shosse 31, 115409 Moscow, Russia}
\author{A.~D.~Panferov}
\author{S.~A.~Smolyansky}
\affiliation{Department of Physics, Saratov State University, 410071
Saratov, Russia}

\date{\today}

\begin{abstract}
The assistance of an intense optical laser pulse on electron-positron pair
production by the Breit-Wheeler and Schwinger processes in XFEL fields is
analyzed.
The impact of a laser beam on high-energy photon collisions with XFEL photons
consists in a phase space redistribution of the pairs emerging in the
Breit-Wheeler sub-process.
We provide numerical examples of the differential cross section for parameters
related to the European XFEL.
Analogously, the Schwinger type pair production in pulsed fields with
oscillating components referring to a superposition of optical laser and XFEL
frequencies is evaluated.
The residual phase space distribution of created pairs is sensitive to the
pulse shape and may differ significantly from transiently achieved mode
occupations.
\end{abstract}


\maketitle

\section{Introduction}

The growing availability of x-ray free electron lasers (XFELs) worldwide
stimulates rethinking of elementary quantum processes in which pairs of
particles and anti-particles, e.g.\ electrons ($e^-$) and positrons ($e^+$), are
created.
An avenue to pair creation is the conversion of light ($\gamma$) into matter in
the collision of (high energy) photon beams.
The Breit-Wheeler process, for instance, is the reaction
$\gamma^\prime + \gamma \to e^+ + e^-$, being a crossing channel of the Compton
process or the time-reversed annihilation.
The famous experiment E-144~\cite{burke_positron_1997} can be interpreted as a
two-step process with (i) Compton backscattering of an optical laser off the
SLAC electron beam and (ii) subsequent reaction of the high-energy
Compton-backscattered photons with the same laser beam, producing the pair.
While the complete sequence of reactions is named trident process, step (ii)
refers to Breit-Wheeler pair production.
The notion non-linear Breit-Wheeler process means the instantaneous reaction
with a multiple of laser beam photons.
A particular variant thereof is the laser assisted Breit-Wheeler process
$\gamma^\prime + \gamma_{XFEL} + \gamma_L \to e^+ + e^-$, i.e.\ the pair
creation in the collision of a probe photon $\gamma_{X^\prime}$ with
co-propagating XFEL ($\gamma_{XFEL}$) and laser ($\gamma_L$) beams.

In contrast to pair creation in counter propagating null fields, also other
electromagnetic fields qualify for pair production.
An outstanding example is the Schwinger effect originally meaning the
instability of a spatially homogeneous, purely electric field with respect to
the decay into a state with pairs and a screened electric field~\cite{schwinger}
(cf.~\cite{gelis_schwinger_2015} for a recent review).
The pair creation rate $\propto \exp\{ - \pi E_c / \vert \vec E \vert \}$ for
electric fields fields $\vec E$ attainable presently in mesoscopic laboratory
installations is exceedingly small since the Sauter-Schwinger (critical) field
strength $E_c = m^2/|e| = \SI{1.3e16}{V/cm}$ for electrons/positrons with masses
$m$ and charges $\pm e$ is so large (we employ here natural units with
$c = \hbar = 1$).
Since the Coulomb fields accompanying heavy and super-heavy atomic nuclei or
ions in a near-by passage can achieve ${\cal O} (E_c)$, the vacuum break down
for such configurations with inhomogeneous static or slowly varying fields have
been studied extensively~\cite{greiner_3,rafelski_superheavy_1971,
muller_solution_1972,muller_solution_1973,fillion-gourdeau_enhanced_2013}.
Experiments, however, were not yet conclusive.

An analogous situation is meet where a spatially homogeneous electric field has
a time dependence.
The particular case of a periodic field is dealt with in~\cite{brezin_pair_1970}
with the motivation that tightly focused laser beams can provide high field
strengths.
The superposition of a few laser beams, as considered, e.g.\ in~\cite
{narozhny_pair_2004}, can enlarge the pair yield noticeably.
A particular variant is the superposition of strong optical laser beams and
weaker but high-frequency (XFEL) beams.
If the frequency of the first field is negligibly small while that of the second
field is sufficiently large, the tunneling path through the positron-electron
gap is shortened by the assistance of the multi-photon effect and, as a
consequence, the pair production is enhanced.
This is the dynamically assisted Schwinger process~\cite
{schutzhold_dynamically_2008}.
As assisted dynamical Schwinger effect one can denote the pair creation (vacuum
decay) where the time dependence of both fields matters.
Many investigations in this context are constrained to spatially homogeneous
field models, that is to the homogeneity region of anti-nodes of pairwise
counter propagating and suitably polarized beams.
Accounting for spatial gradients is much more challenging~\cite
{dunne_worldline_2005,ruf_pair_2009}.

Other field combinations, e.g.\ the nuclear Coulomb field and XFEL/laser beams,
are also conceivable~\cite{augustin_nonlinear_2014,di_piazza_effect_2010}, but
will not be addressed here (cf.~\cite{di_piazza_extremely_2012} for a survey).

A few of the above quoted field configurations share as a common feature the
pair creation in bi-frequent fields, as provided by the superposition of optical
laser and XFEL beams.
Accordingly, we are going to consider the laser assisted Breit-Wheeler and
dynamical Schwinger processes in such bi-frequent fields.
Our paper is organized as follows.
Section 2 deals with the laser assisted Breit-Wheeler process, where spectral
caustics have been identified already in~\cite{nousch_spectral_2016,
otto_pair_2016}.
Specifically, we deliver here as new result the phase space distribution of
positrons, in particular the double-differential cross section as a function of
longitudinal and transverse momenta.
In section 3 we consider the assisted dynamical Schwinger effect for the
superposition of two spatially homogeneous fields of different strengths and
frequencies with a common pulse envelope, as investigated in~\cite
{otto_lifting_2015,otto_dynamical_2015,panferov_assisted_2016,otto_pair_2016}.
Here we present for the first time examples of the time evolution to show the
striking difference of the transient mode occupancy in an adiabatic basis and
the residual phase space yield.
The summary and discussion can be found in section 4.

\section{Laser assisted Breit-Wheeler process}

The laser assisted, non-linear Breit-Wheeler process is dealt with within the
strong-field QED (Furry picture) as decay of a probe photon traveling through a
null field $A$, symbolically $\gamma' \to e^+_A + e^-_A$ where $e^\pm_A$
denote Volkov solutions of the Dirac equation in a plane wave with vector
potential
\begin{equation}
A^\mu (\phi) = \hat a_X f_X (\phi) \varepsilon^\mu_X \cos \phi
+ \hat a_L f_L (\eta \phi) \varepsilon^\mu_L \cos \eta \phi.
\label{N1}
\end{equation}
The field~\eqref{N1} is a classical background field, while the probe photon
belongs to a quantized radiation field.
The XFEL (frequency $\omega_X$, four-momentum $k^\mu_X$, intensity parameter
$a_0^{(X)} = \hat a_X|e|/m$, polarization four-vector $\varepsilon^\mu_X$) and
laser (frequency $\omega_L = \eta \omega_X$, intensity parameter
$a_0^{(L)} = \hat a_L|e|/m$, polarization four-vector $\varepsilon^\mu_L$) beams are co-propagating and their linear polarizations are perpendicular to each
other.
Both beams are pulsed as described by the envelope functions
$f_X = \exp\{ - \phi^2 /(2 \tau_X^2)\}$ and
$f_L = \cos^2 \left(\pi \phi /(2 \tau_L) \right)$ for
$-\tau_L \le \phi \le + \tau_L$ and zero elsewhere.
The invariant phase is $\phi = k_X\!\cdot\!x$ with a dot indicating the scalar
product of the four-wave vector $k_X$ and the space-time coordinate $x$.

The theoretical basis for formulating and evaluating the cross section (as well
as the corresponding kinematics) is described in Ref.~\cite
{nousch_spectral_2016}.
It reads:
\begin{equation}
\frac{\diff \sigma}{\diff p_\perp \diff p_\parallel \diff\varphi} =
\frac{e^2 p_\perp}{(4\pi)^3p_0\rho_Xj_{in}\,k_X\!\cdot\!(k_{X'}-p)}
\sum_{spins} |\mathcal M|^2
\label{N2}
\end{equation}
with matrix element ${\cal M} = \int\!\diff^4 x\, \bar\Psi_A
\slashed\varepsilon_{X'} \exp\{ik_{X'}\!\cdot\!x\} \Psi_A$,
$\Psi_A$ is the Volkov solution in the external classical field $A$
from~\eqref{N1} and $\bar \Psi_A$ its adjoint, and  $\varepsilon_{X'}$ denotes
the four-polarization of the probe photon $X'$ (four-momentum $k_{X'}$) which
will be averaged.
We normalize with the particle density $\rho_X = m^2a_{X}^2 /(2e^2)
\int_{-\infty}^\infty \diff\phi f^{2}_X(\phi)$ and by the incoming photon flux
$j_{in}= k_X\!\cdot\!k_{X'} / (k_X^0 k_{X'}^0)$ such that without the laser
assistance, Eq.~\eqref{N2} recovers the standard Breit-Wheeler cross section.
Examples for the transverse momentum ($p_\perp$) distribution of positrons are
presented in~\cite{nousch_spectral_2016,otto_pair_2016} for selected values of
the longitudinal momentum $p_\parallel$ at azimuthal angle $\varphi = \pi$
measured w.r.t.\ the laser polarization plane.\footnote{$p_\parallel$ is
parallel to the laser plane and $p_\perp$ perpendicular to it.}

\begin{figure}
\def\w{0.49}
\centering
\includegraphics[width=\w\textwidth]{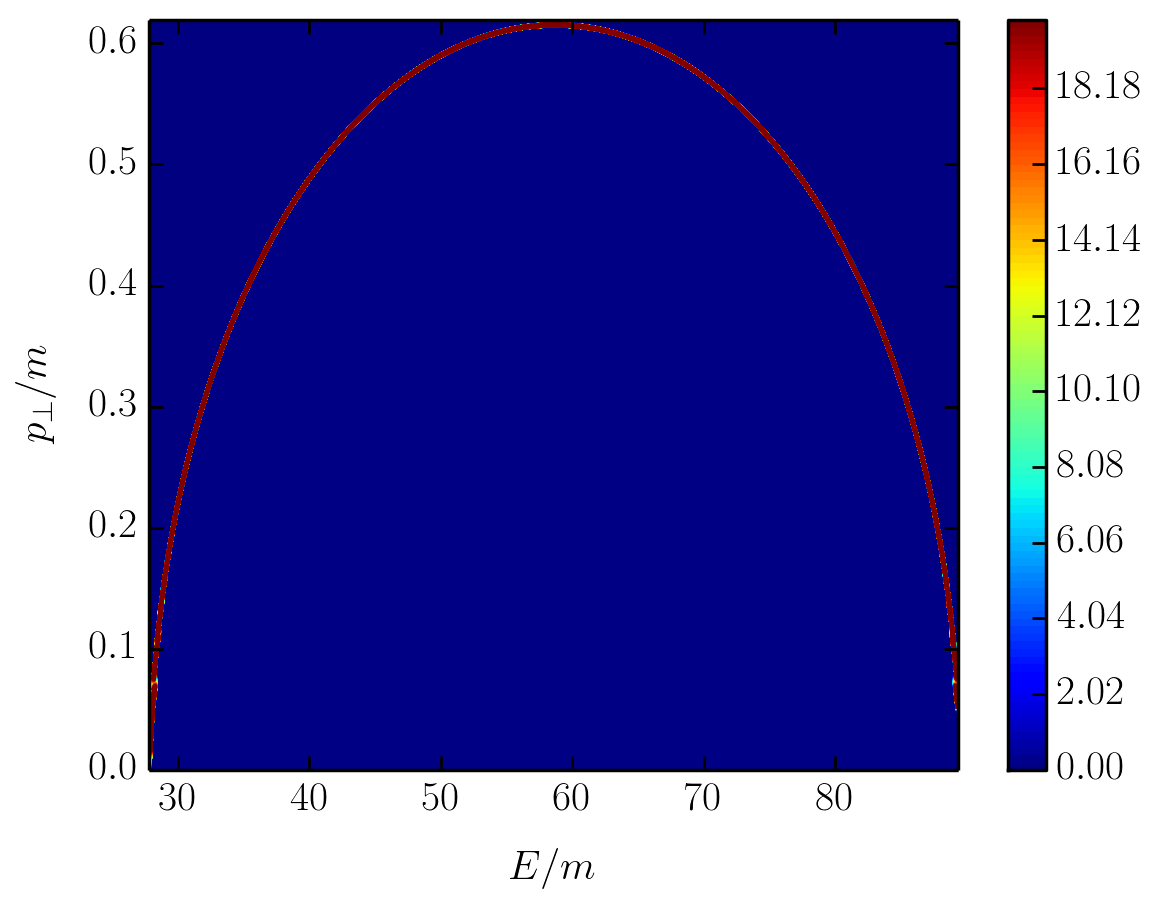}\hfill
\includegraphics[width=\w\textwidth]{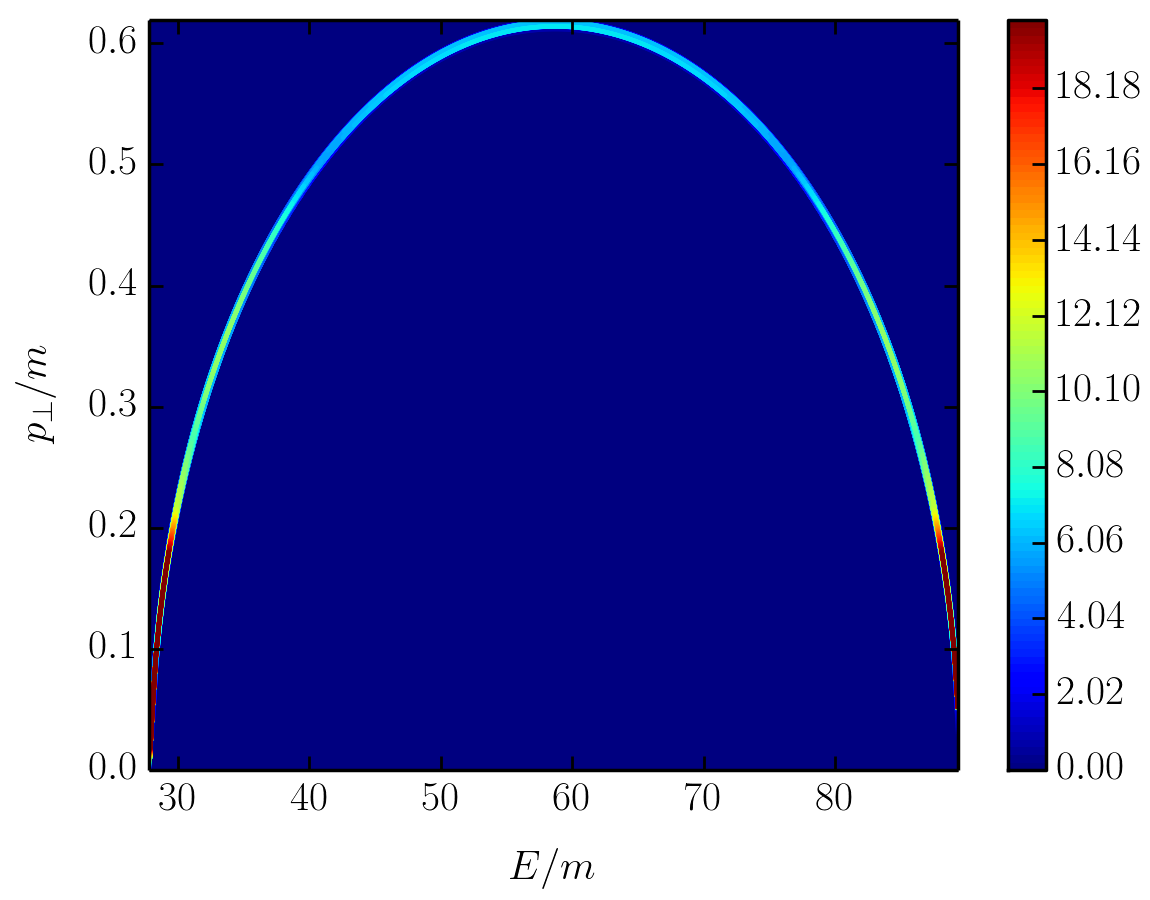}\\
\includegraphics[width=\w\textwidth]{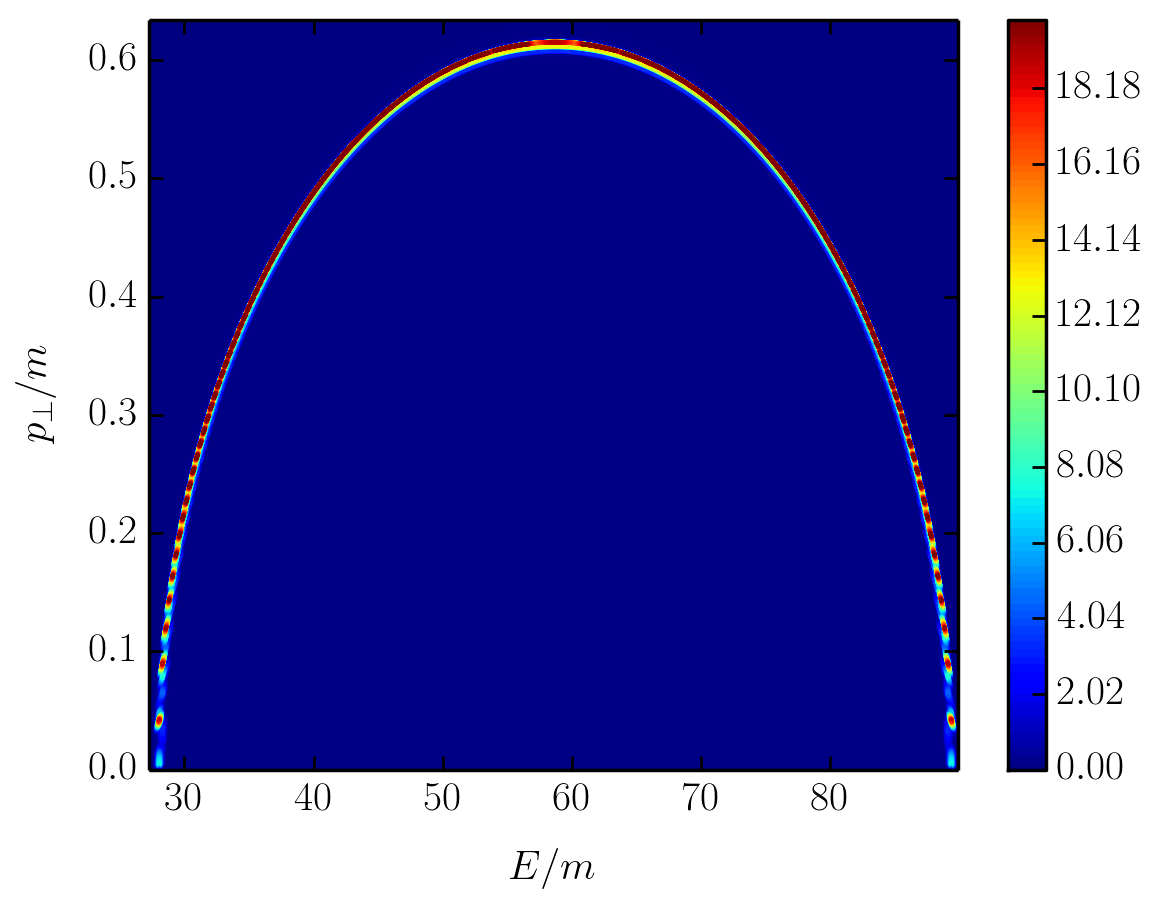}\hfill
\includegraphics[width=\w\textwidth]{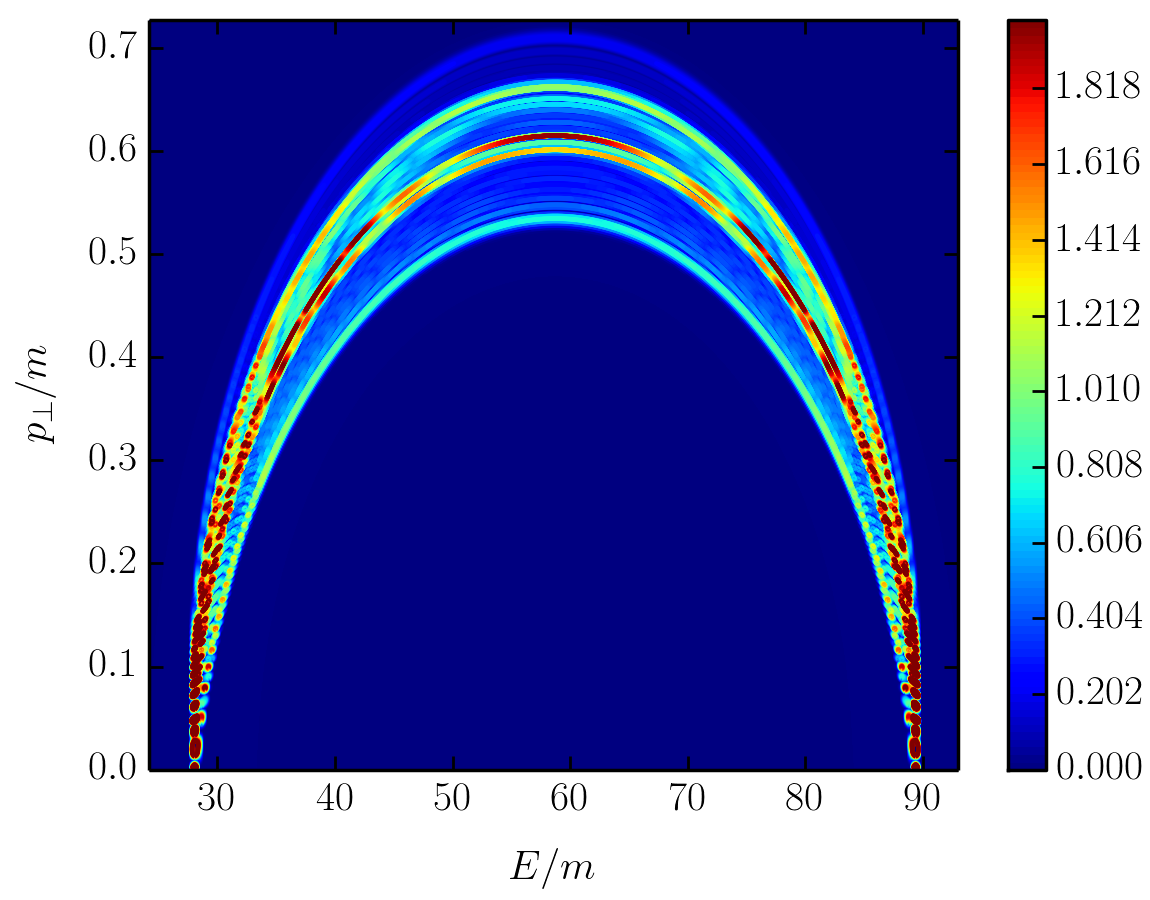}\\
\caption{
Color-contour plots of the phase space distribution of positrons in a plane
aligned to the laser polarization by $\varphi=0.5\pi$ (left panels) and in the
laser polarization plane, i.e. at $\varphi=\pi$ (right panels), as well as
$a_0^{(L)}=0.01$ (upper panels) and $a_0^{(L)}=0.1$ (lower panels).
Transverse momentum $p_\perp$ and energy $E = (m^2 + p_\perp^2 + p_\parallel^2
)^\frac12$ are scaled by the electron mass $m$.
Parameters: $\omega_{X'}=\SI{60}{MeV}$, $\omega_X=\SI{6}{keV}$,
$\omega_L=\SI{10}{eV}$, $\tau_L=4\pi$, $\tau_X=7/\eta$.
Note that $a_{0}^{(X)}$ does not enter the cross section since we consider here
the leading order contribution in an expansion in powers of $a_{0}^{(X)}\ll1$,
which applies for present XFEL facilities~\cite{ringwald_pair_2001}.
}
\label{fig1N}
\end{figure}

\begin{figure}
\def\w{0.49}
\centering
\includegraphics[width=\w\textwidth]{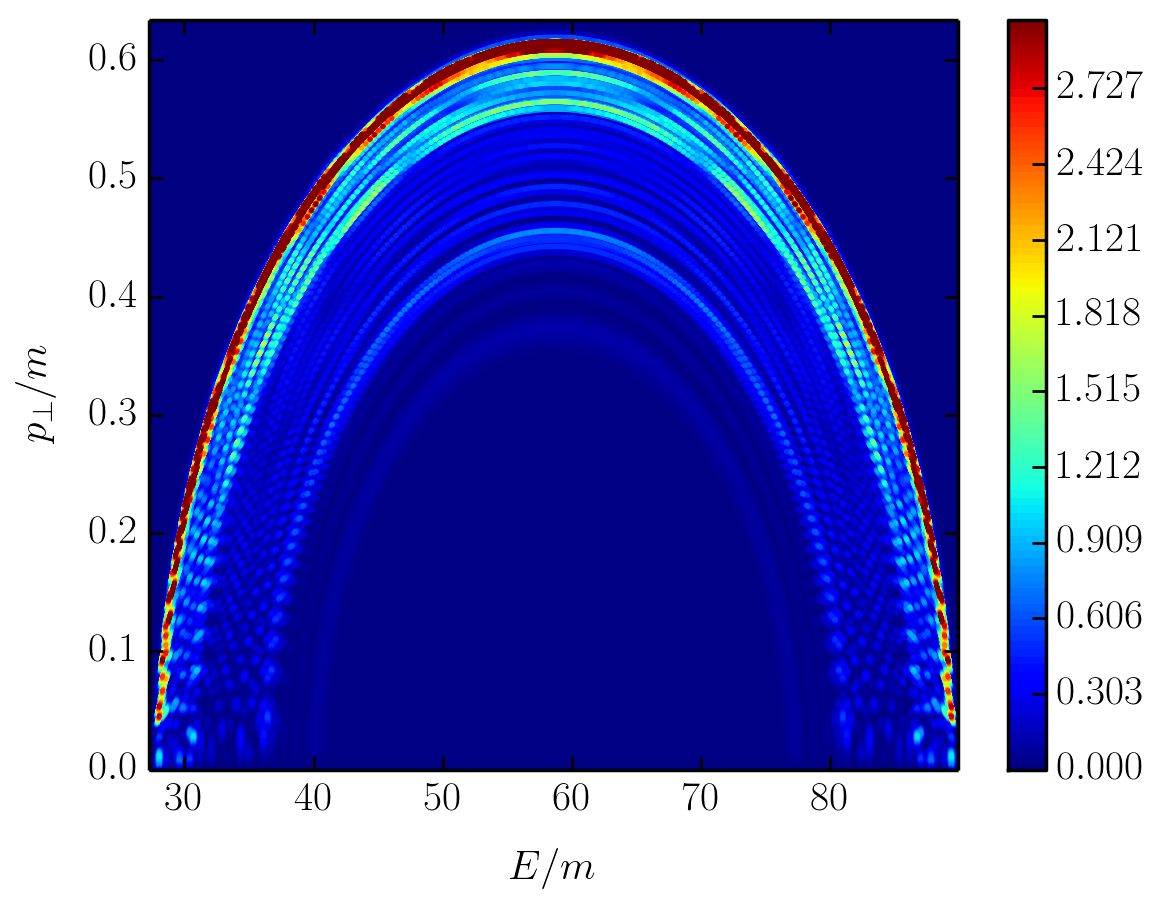}\hfill
\includegraphics[width=\w\textwidth]{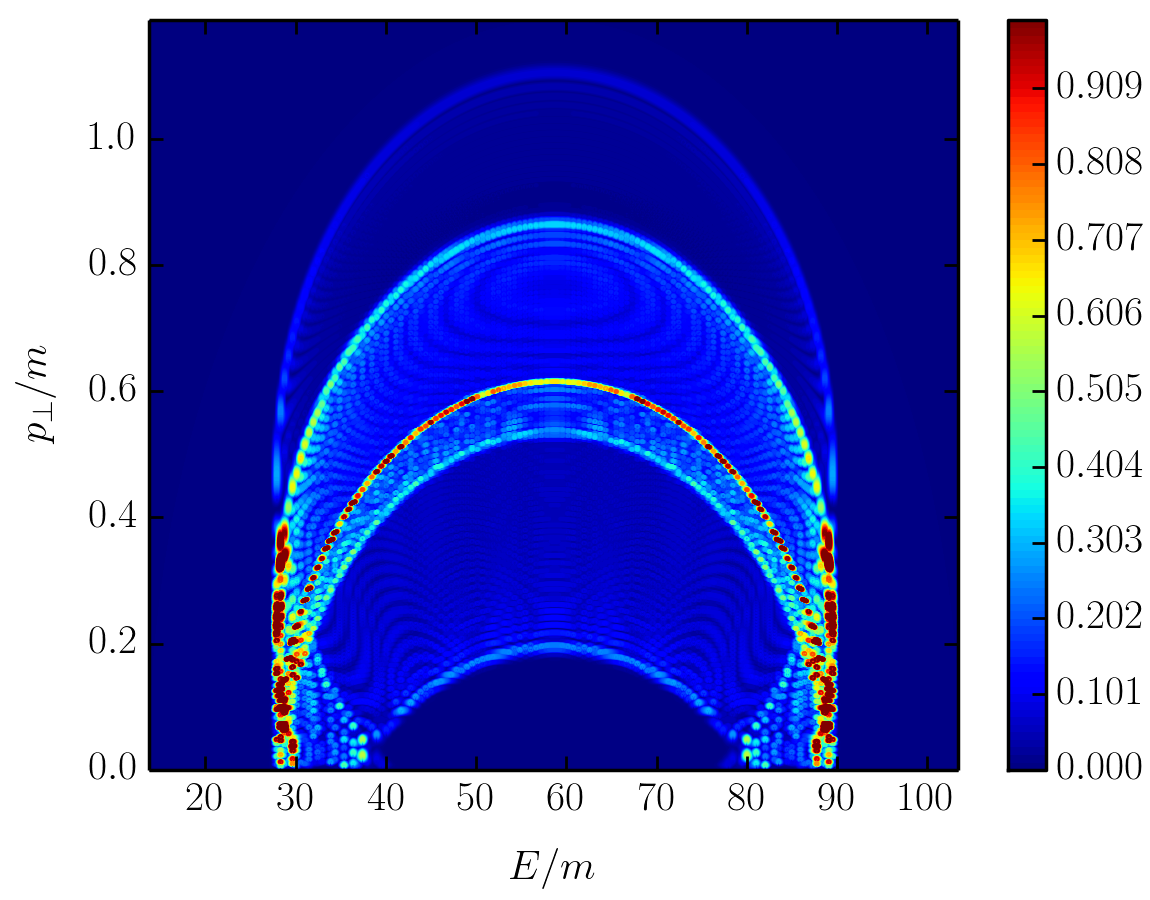}\\
\includegraphics[width=\w\textwidth]{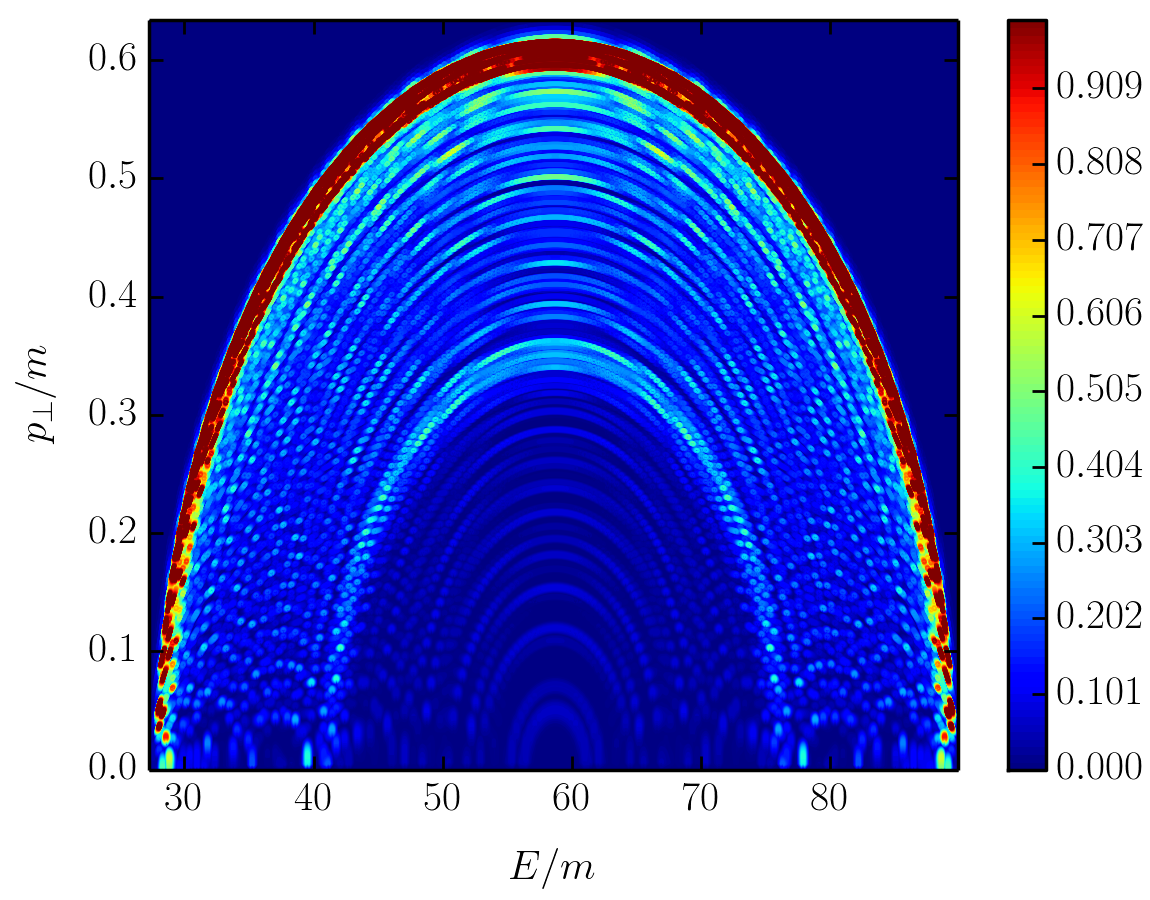}\hfill
\includegraphics[width=\w\textwidth]{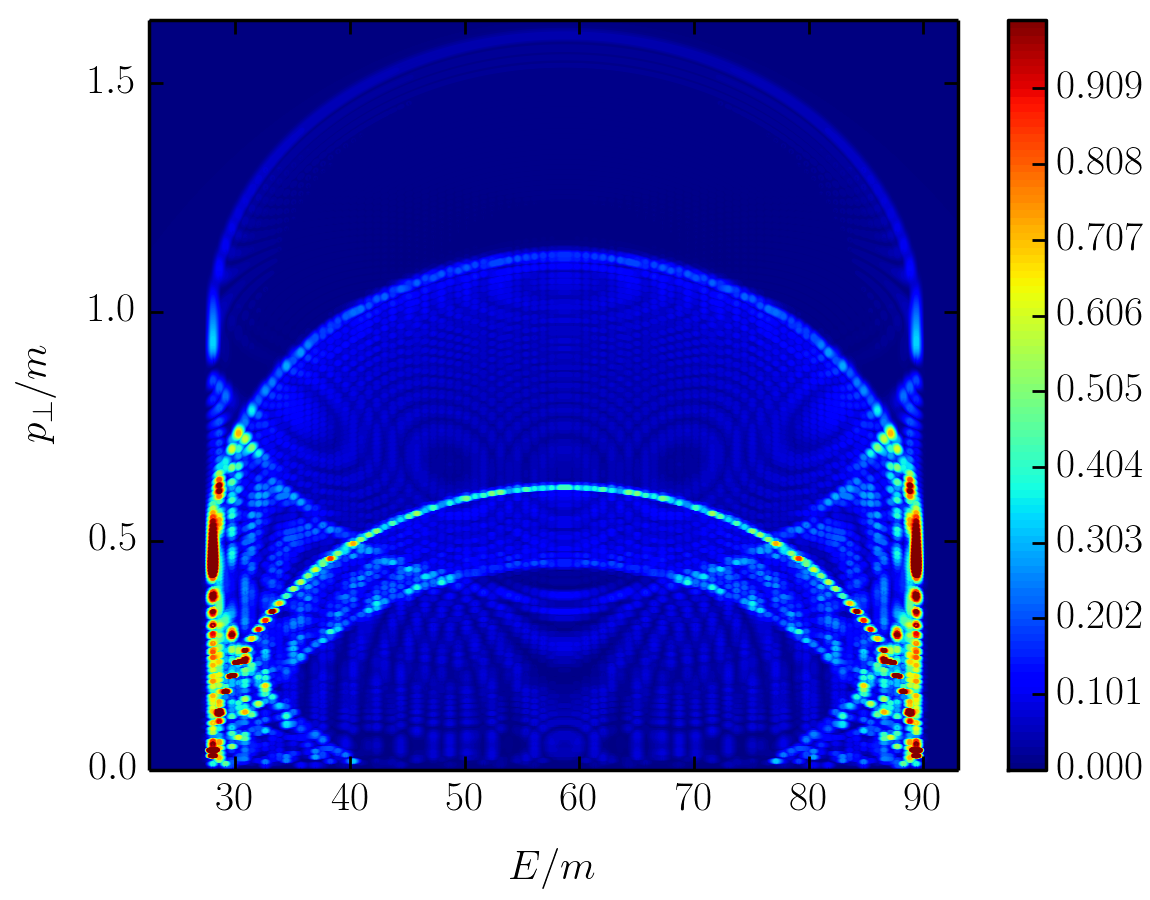}\\
\caption{
Same as Fig.~\ref{fig1N} but for $a_{0}^{(L)}=0.5$ (upper panels) and
$a_{0}^{(L)}=1$ (lower panels).
}
\label{fig2N}
\end{figure}

To complete the information on the phase space distribution we display the
double-differential cross section $\diff\sigma / \diff p_\perp \diff p_\parallel
\diff\varphi$ at $\varphi = \pi/2$ and $\pi$ as a contour plot over the
$p_\perp - p_\parallel$ plane, see Figs.~\ref{fig1N} and~\ref{fig2N}.
The ridges as loci of accumulated intensity are interpreted in line with~\cite
{seipt_caustic_2016} as spectral caustics related to stationary phase points.
The impact of the laser consists of a redistribution of Breit-Wheeler-produced
pairs in the phase space.
Without the laser (this means $\hat a_L = 0$) the spectrum becomes much simpler
and squeezed to a narrow region (see upper row in Fig.~\ref{fig1N} for
$a_O^{(L)}=0.01$), and only the finite pulse length $\tau_X$ has imprints on the
spectral distribution~\cite{nousch_pair_2012,titov_enhanced_2012,
titov_breit-wheeler_2013}.
With increasing laser intensity, parametrized by $a_0^{(L)}$ or $\hat a_L$, the
spectra become stretched, both in $p_\perp$ and $p_\parallel$ (respective energy
$E$) directions, see bottom row in Fig.~\ref{fig1N} and both rows in Fig.~\ref
{fig2N}.
The effect of the laser assistance is strongest in the polarization plane of the
laser for moderate intensities.
Due to the Lorentz force, at larger intensities, also the off-plane becomes
populated, see left columns in Fig.~\ref{fig2N}.


\section{Assisted dynamical Schwinger process}
\begin{figure}
\centering
\includegraphics[width=0.97\textwidth]{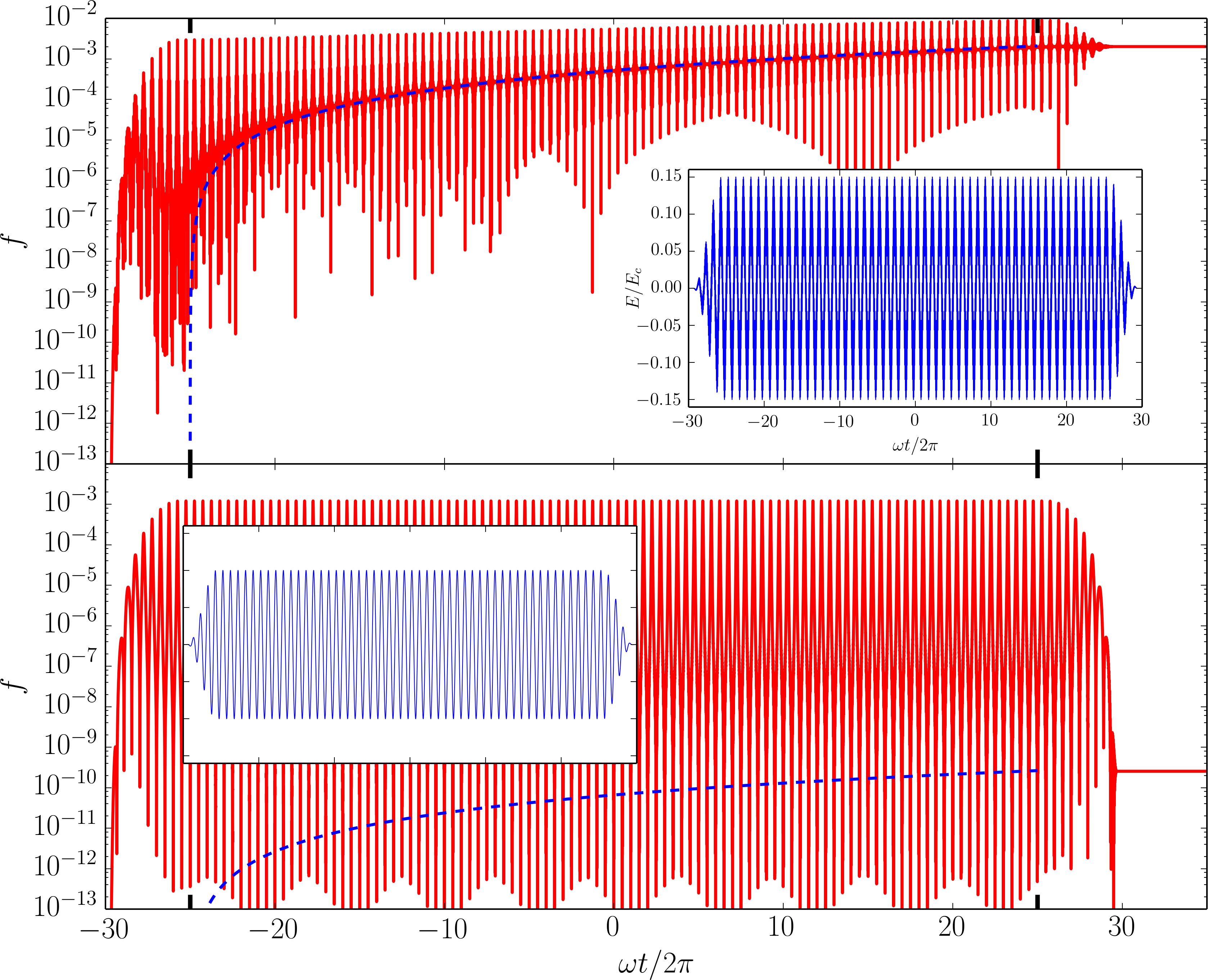}
\caption{Time evolution of $f(p_\perp, p_\parallel, t)$ (solid curves, note the
variation over many orders of magnitude) from the full quantum kinetic
equation~\eqref{O1} for the same envelope function $K$ in~\eqref{O2} as in~\cite
{otto_lifting_2015,otto_dynamical_2015}.
The dashed curves in the flat-top interval $-t_\text{f.t.}/2 < t <
t_\text{f.t.}/2$ (marked by bold ticks) are for the relevant component
$f_\text{rel}$ of $f$ (defined in~(\ref{f_rel1}-\ref{f_rel3})) which becomes
asymptotically the residual yield.
Note that $\diff f/\diff t = 0$ for $t > t_\text{f.t.}/2 + t_\text{ramp}$
according to~\eqref{O1} since the external field vanishes.
The insets display the time structure of the electric fields.
For $\omega t_\text{ramp}=5\cdot2\pi$, $\omega t_\text{f.t.}=50\cdot2\pi$,
$E_1=0.1E_c$, $\omega=0.02m$, $p_\parallel=0$.
Further parameters are $E_2=0.05E_c$, $N=25$, $p_\perp=\num{0.155325}m$ (upper
panel) and $E_2=0$, $p_\perp=\num{0.161900}m$ (lower panel, the same inset
labels and axes ranges as in upper panel inset).
}
\label{fig1O}
\end{figure}
Let us now discuss the time-evolution of the assisted Schwinger pair-production
process in bi-frequent laser pulses.
The quantum kinetic equation~\cite{schmidt_quantum_1998}
\begin{equation}\label{O1}
\frac{\diff}{\diff t} f(\vec p ,t) = \frac 12 \lambda(\vec p, t )
\int\limits^t_{- \infty}\!\! \diff t'
\lambda(\vec{p}, t') (1 - 2 f(\vec{p}, t'))
\cos\theta(\vec{p}, t, t')
\end{equation}
determines the time ($t$) evolution of the dimensionless phase space
distribution function per spin projection degree of freedom
$f(\vec p, t) = \diff^6 N(\vec p, t) / \diff^3 p\, \diff^3 x$ from a vacuum
state $f(\vec p, t \to -\infty) = 0$.
Here, the quantities $\lambda(\vec p,t) = eE(t) \, \varepsilon_\perp(p_\perp)
\varepsilon^{-2}(\vec{p}, t)$ stand for the vacuum transition amplitude, and
$\theta(\vec p, t, t') = 2 \int^t_{t'} \diff\tau \, \varepsilon(\vec p, \tau)$
for the dynamical phase;
the quasi-energy $\varepsilon$, the transverse energy $\varepsilon_\perp$ and
the longitudinal quasi-momentum $P$ are defined by
$\varepsilon(\vec p, t) = \sqrt{\varepsilon^2(p_\perp) + P^2(p_\parallel, t)}$,
and $\varepsilon_\perp(p_\perp)= \sqrt{m^2 + p^2_\bot}$,
$P(p_\parallel, t) = p_\parallel -eA(t)$, where $p_\perp=|\vec p_\perp|$ is the
modulus of the momentum component perpendicular to the electric field, and
$p_\parallel$ denotes the $\vec E$-parallel momentum component.
The electric field $E = -\dot A$ in Coulomb gauge follows from the potential
model
\begin{equation}
A = K(\omega t) \left( \frac{E_1}{\omega} \cos (\omega t)
+ \frac{E_2}{N \omega} \cos (N \omega t) \right).
\label{O2}
\end{equation}
Equation (\ref{O2}) describes again a classical, spatially homogeneous,
bi-frequent field with frequency ratio $N$ (integer) and field strengths $E_1$
-- the strong field ``1'' -- and $E_2$ -- the weak field ``2''.
The quantitiy $K$ is the common envelope function with the properties (i) flat
in the flat-top time interval $-t_{f.t.} / 2 < t < + t_{f.t.}/2$ and (ii) zero
for $t < - t_{f.t.}/2 - t_{ramp}$ and $t > t_{f.t.}/2 + t_{ramp}$ and (iii)
smooth everywhere, i.e.\ $K$ belongs to the $C^\infty$ class;
$t_{ramp}$ is the ramping duration characterizing the switching on/off time
intervals (see~\cite{otto_lifting_2015,otto_dynamical_2015} for details;
other envelopes are dealt with in~\cite{panferov_assisted_2016};
carrier envelope phase effects and further effects of different envelope models
deserve further dedicated investigations).
We emphasize the unavoidable ambiguity of a particle definition at intermediate
times~\cite{dabrowski_super-adiabatic_2014}, i.e.\ only
$f(p_\perp, p_\parallel, t \to +\infty)$ can be considered as a single particle
distribution which may represent the source term of a subsequent time evolution.
Screening and back reaction need not to be included for small values of $f$.

Examples of the residual phase space distribution
$f(p_\perp, p_\parallel, t \to + \infty)$ can be found in~\cite
{otto_lifting_2015,otto_dynamical_2015}.
In essence, for large enough values of $N$, the field $E_2$ enhances the yield
achievable by the field $E_1$ alone.
The enhancement can be gigantic, but field strengths $E_1$ in the order of such
ones envisages in ELI-IV~\cite{eli,eli-np} and beyond HiPER~\cite{hiper} and
sufficiently large $N$ are required to overcome the exponential suppression of
pair production in sub-critical fields.
The enhancement by a second, high-frequency field is in agreement with a general
statement in~\cite{ilderton_nonperturbative_2015}:
The pair production probability is increased by temporal inhomogeneities.

It is instructive to inspect the approach to the residual distribution
$f(p_\perp, p_\parallel, t \to + \infty)$ by means of Eq.~(\ref{O1}).
Figure \ref{fig1O} exhibits examples of the time evolution of $f$ (solid curves)
in two phase space points $p_\perp = p_\ell$, $p_\parallel = 0$ where a
resonance condition (cf.~\cite{otto_dynamical_2015} for the definition of a
series of $p_\ell$ values) is fulfilled.
The upper panel is for a bi-frequent field, while the lower panel is for a
single field.
Note the large difference of the residual phase space occupancy upon the
assistance of a weak but fast field ``2''.
There are rapid oscillations with huge maximum values at transient times which,
however, drop significantly upon switching off the external field.
In~\cite{otto_dynamical_2015} an approximation has been presented which allows
to follow a particularly relevant component of $f$, $f_\text{rel}$ (dashed
curves), which becomes the residual yield at $t \to \infty$:
\begin{align}
&f(p_\ell, 0, t) \approx f_\text{osc}(p_\ell, 0, t)
+ f_\text{rel}(p_\ell, 0, t)\:,\label{f_rel1}\\
&f_\text{rel}(p_\ell, 0, t) = \frac{1}{2} |F_\ell|^2 t^2\:,\label{f_rel2}\\
&F_\ell = \frac{\omega}{2\pi}\!\! \int\limits_0^{2\pi/\omega}\!\!\diff t\,
\lambda(p_\ell, 0, t)\mathrm e^{i\theta(p_\ell, 0, t, 0)}\:,\label{f_rel3}
\end{align}
where $f_\text{osc}$ refers to the irrelevant oscillating part and $F_\ell$ is a
Fourier coefficient in the low-density approximation (cf.~\cite
{otto_lifting_2015}).
In fact, the dashed curves give a remarkably accurate estimate of the final
value, irrespectively of details of the ramping and de-ramping as long as the
slowly varying envelope approximation is applicable and $t_{f.t.} \gg t_{ramp}$.
Large differences of the residual yields in neighboring phase space points point
to resonance type structures.

\section{Summary and Discussion}

In summary we have extended our previous studies~\cite{otto_lifting_2015,
otto_dynamical_2015,panferov_assisted_2016} and deliver here further important
details of (i) the phase space distribution in the laser assisted Breit-Wheeler
process and (ii) the time evolution of the mode occupancy in the assisted
dynamical Schwinger effect.
Both topics are motivated by the availability of x-rays by XFELs and upcoming
ulta-high intensity laser beams.
We consider the perspectives offered by the combination of both beam types
resulting in bi-frequent fields.

The laser assisted Breit-Wheeler process is studied for the head-on collision of
a probe photon beam with two co-propagating beams, provided by an optical laser
(L) and an XFEL (X).
Despite of the coherence of the XFEL beam, its intensity parameter $a_0^{(X)}$
is small, thus calling for a restriction of leading order effects in powers of
$a_0^{(X)}$.
The treatment of misalignment effects of the L and XFEL beams as well as higher
order effects in $a_0^{(X)}$ is left for future work, as the investigations of
realistic focal spot geometries in focused beams and general polarization
effects as well as carrier envelope phase effects.
The beams considered here are represented by null fields with large frequency
ratios: $\omega_{X^\prime} = \cal O(\SI{60}{MeV})$,
$\omega_X = \cal O(\SI{6}{keV})$ and $\omega_L = \cal O(\SI{10}{eV})$ in the
laboratory.
The impact of the laser L consists essentially in a reshuffeling of the phase
space distribution for the considered parameters.

Our analysis of the dynamical Schwinger process is based on a very special
background field model assuming that spatial inhomogeneities can be neglected.
Assuming further that pair production happens in a spatial region of the
dimension of the electron Compton wave length, the often posed idea refers to
such a small region in the anti-nodes of a standing wave created by counter
propagating and suitably polarized laser beams, where essentially an oscillating
electric field occurs.
Having in mind the principal interest in the Schwinger effect as genuinely
non-perturbative quantum decay of the vacuum, we stay with such a model and
extend it to a bi-frequent field.
To overcome the exponential suppression of pair creation one has to combine a
near-critical, low-frequency (laser) field and a sub-critical, high-frequency
field, the latter one corresponding more to $\gamma$ radiation than x-rays.
We expect that deviations from the considered idealization will diminish the
pair yield, despite of potentially huge enhancement effects due to the
assistance of a second field.
An interesting question concerns the speculation whether the fairly large
transient mode occupation can be probed, e.g.\ via secondary signals or active
probes.

\section*{Acknowledgments}

R.~Sauerbrey, T.~E.~Cowan and H.~Takabe are gratefully acknowledged for the
collaboration within the HIBEF project~\cite{hibef}.
We thank S.~Fritzsche and A.~Surzhykov for the common work on the caustic
interpretation of elementary QED processes in bi-frequent fields.
D.B. and S.A.S have been supported by Narodowe Centrum Nauki under grant number
UMO-2014/15/B/ST2/03752.\bigskip

We dedicate this article to Walter Greiner on the occasion of his 80th birthday.
Walter Greiner promoted essentially the in-depth exploration of the nature of
the quantum vacuum.

\bibliographystyle{apsrev4-1}
\bibliography{lit}
\end{document}